\documentclass[10pt,aps,prx,twocolumn,superscriptaddress,floatfix]{revtex4-2}
\usepackage{braket}
\usepackage{amsfonts, bm, mathtools}
\usepackage{booktabs}
\usepackage{siunitx}
\usepackage{xcolor}
\usepackage{xurl}
\usepackage[breaklinks=true,colorlinks,citecolor=blue,linkcolor=blue,urlcolor=blue]{hyperref}
\usepackage{txfonts}
\usepackage{orcidlink}
\usepackage{float}
\usepackage{algorithm}
\usepackage{algorithmic}



\begin{document}

\title{Active Sampling Sample-based Quantum Diagonalization from Finite-Shot Measurements}

\author{Rinka Miura$^{\orcidlink{0009-0003-6502-2243}}$}
\email[]{acdccit5@gmail.com}
\affiliation{Department of Applied Chemistry, Kobe City College of Technology, Japan}

\date{\today}

\begin{abstract}
Near-term quantum devices provide only finite-shot computational-basis measurement outcomes and typically prepare imperfect, contaminated states rather than exact ground states. This practical constraint motivates algorithms that can convert samples into reliable estimates of low-energy properties without full state tomography or exhaustive Hamiltonian measurement. In this work we propose Active Sampling Sample-based Quantum Diagonalization (AS-SQD), an approach that frames Sample-based Quantum Diagonalization (SQD) as an active learning problem: given a finite multiset of measured bitstrings, which additional basis states should be included in the effective subspace to most efficiently recover the true ground-state energy? SQD constructs an effective low-dimensional eigenvalue problem by restricting the Hamiltonian to the span of a selected set of computational basis states and then classically diagonalizing the restricted matrix. However, naive SQD that only uses the sampled subspace often suffers from severe bias under finite-shot sampling and excited-state contamination, while blind subspace expansion (e.g., random exploration of connected basis states) is inefficient and unstable as system size grows. We introduce a perturbation-theoretic acquisition function based on Epstein--Nesbet second-order energy corrections to rank candidate basis states that are connected to the current subspace by the Hamiltonian. At each iteration, AS-SQD (i) diagonalizes the restricted Hamiltonian to obtain an approximate ground state, (ii) generates a candidate set of connected basis states, and (iii) adds the most valuable candidates according to a scoring function derived from perturbation theory. We evaluate AS-SQD on disordered Heisenberg and Transverse-Field Ising (TFIM) spin chains up to 16 qubits under a realistic preparation model that mixes 80\% ground state and 20\% first excited state. Furthermore, we validate the inherent robustness of our approach against real-world state preparation and measurement (SPAM) errors using physical samples directly from an IBM Quantum processor. Across both simulated and physical evaluations, AS-SQD consistently achieves substantially lower median absolute energy errors than standard SQD and random expansion. Detailed ablation studies isolate the driving mechanisms of the perturbation score, demonstrating that physics-guided basis acquisition effectively concentrates computation on energetically relevant directions and bypasses exponential combinatorial bottlenecks.
\end{abstract}

\maketitle

\section{Introduction}
Noisy intermediate-scale quantum (NISQ) devices promise access to quantum dynamics and many-body properties beyond the reach of exact classical simulation, yet they impose stringent limitations: (i) measurements yield only finite-shot samples in a chosen basis, (ii) prepared states are imperfect and often contain excited-state contamination, and (iii) full Hamiltonian matrix access is unavailable \cite{preskill2018nisq,cerezo2021variational}. As a result, many practical quantum algorithms reduce to the question of how to best convert limited measurement data into accurate physical predictions.

A core target task is estimating the ground-state energy $E_0$ of a Hamiltonian $H$ expressed as a sum of Pauli strings, $H = \sum_\ell c_\ell P_\ell$, where each $P_\ell$ is a tensor product of single-qubit Pauli operators. The variational quantum eigensolver (VQE) is a standard approach: prepare a parametrized quantum state and estimate energy by measuring each term of $H$ \cite{peruzzo2014vqe,kandala2017hardware}. However, VQE can be limited by shot complexity, optimization difficulties, and errors in state preparation and measurement (SPAM)\cite{mcclean2018barren}. Moreover, even if a quantum device produces a state close to the ground state, we do not directly obtain its amplitude vector $\psi$; we only observe samples from its measurement distribution in a basis.

This motivates a complementary family of \emph{subspace} approaches that use a restricted basis to represent and diagonalize an effective Hamiltonian. One classical analogue is selected configuration interaction (selected CI), where a small set of determinants is iteratively expanded using perturbative importance measures (e.g., CIPSI, ASCI) \cite{huron1973cipsi,tubman2016asci,holmes2016hci}. On the quantum side, the Quantum Subspace Expansion (QSE) expands around a reference state using excitation operators and measures matrix elements in that subspace \cite{mcclean2017qse}. These methods suggest that adaptive, physics-motivated subspace selection can overcome the combinatorial dimension of the Hilbert space.

In this paper, we focus on a particularly constrained regime: \emph{we assume the solver has access to (a) the Hamiltonian in Pauli-string form and (b) only computational-basis measurement outcomes (bitstrings) sampled from an imperfectly prepared quantum state.} From these bitstrings alone, we build a low-dimensional subspace spanned by the observed basis states and then classically diagonalize the Hamiltonian restricted to that subspace. We refer to this procedure as \emph{Sample-based Quantum Diagonalization (SQD)}:
\begin{equation}
E(S) = \min_{\lvert\psi\rangle\in\mathrm{span}(S)}
\frac{\langle\psi\lvert H\rvert\psi\rangle}{\langle\psi\vert\psi\rangle},
\label{eq:restricted_rayleigh}
\end{equation}
where $S$ is a selected set of computational basis states (bitstrings).

While SQD is conceptually simple, it reveals a central algorithmic problem: \emph{Given an initial sample-derived subspace $S$, which additional basis state $k$ should be added to most improve $E(S)$ toward the true ground energy $E_0$?} A purely sample-defined subspace can be too small or biased (especially under contamination), leading to large errors. On the other hand, blindly expanding the subspace by exploring basis states connected by $H$ (e.g., random additions) can require many iterations and can waste compute on irrelevant directions.

Recent advances have further expanded the scope of sample-based quantum diagonalization (SQD) beyond its initial formulation. Symmetry-adapted variants exploit conserved quantities to restrict the effective Hilbert space and improve numerical stability \cite{nogaki2025symm_sqdiagonal}. Krylov-inspired sample-based constructions have been proposed to systematically generate subspaces from measurement data \cite{2501.09702_sb_krylov}, while partitioned quantum subspace expansion techniques address finite-shot noise through structured matrix decompositions \cite{wu2403_partsqse}. Extensions toward quantum chemistry and periodic solid-state systems demonstrate the applicability of SQD-type frameworks beyond small lattice benchmarks \cite{duriez2025_sqdiagonal_solids,piccinelli2025_rand_sqd}. In parallel, adaptive and neural-network-assisted basis selection strategies have begun to explore data-driven subspace optimization \cite{cantori2025_adaptive_nn_sqd}. These developments collectively highlight a growing interest in subspace construction strategies under realistic sampling constraints. However, a principled acquisition rule that directly targets ground-state energy improvement under finite-shot contamination remains largely unexplored.

To address this challenge, we propose \emph{Active Sampling Sample-based Quantum Diagonalization (AS-SQD)}, an adaptive basis acquisition strategy grounded in perturbation theory. AS-SQD casts the subspace growth of SQD as a sequential decision-making problem under finite-shot constraints, selecting the most valuable basis states to add next.  We score candidate basis states using an Epstein--Nesbet partition of the Hilbert space, approximating their expected second-order energy improvement. The method is evaluated on disordered Heisenberg and Transverse-Field Ising Models (TFIM) up to 16 qubits, as well as on IBM Quantum hardware. Our results confirm that AS-SQD achieves higher predictive accuracy than standard SQD and random expansion.

\section{Background}
We consider $n$-qubit Hamiltonians expressed as
\begin{equation}
H = \sum_{\ell=1}^{L} c_\ell P_\ell,\quad
P_\ell \in \{I,X,Y,Z\}^{\otimes n},\quad c_\ell \in \mathbb{R}.
\end{equation}
A Pauli string $P_\ell$ maps computational basis states to (possibly different) basis states up to phases. This implies that matrix elements $\langle b\lvert H\rvert b'\rangle$ can be computed efficiently given the Pauli decomposition, without materializing the full $2^n \times 2^n$ matrix, by applying each Pauli term to a bitstring and accumulating contributions.

In the context of quantum simulation, a quantum device prepares a state $\lvert\psi\rangle$ (possibly noisy) and returns bitstrings $b \in \{0,1\}^n$ sampled from $p(b) = |\langle b\vert\psi\rangle|^2$. With $N_{\mathrm{shots}}$ shots, the observed counts define an empirical distribution $\hat{p}$, which may miss important basis states if their probability is small. Furthermore, prepared states may contain excited-state components. A simple but practically relevant model is a mixture of low-energy eigenstates:
\begin{equation}
p(b) = (1-\eta)|\langle b\vert\psi_0\rangle|^2 + \eta|\langle b\vert\psi_1\rangle|^2,
\label{eq:contamination}
\end{equation}
where $\eta$ is the contamination rate and $\lvert\psi_0\rangle, \lvert\psi_1\rangle$ are the ground and first excited eigenstates.

To extract low-energy properties from these samples, Sample-based Quantum Diagonalization (SQD) utilizes a selected set $S$ of observed basis states. SQD forms the restricted Hamiltonian
\begin{equation}
H_S = \left[\langle s_i\lvert H\rvert s_j\rangle\right]_{i,j=1}^{|S|},
\end{equation}
and solves the eigenvalue problem
\begin{equation}
H_S \bm{c} = E_S \bm{c},
\end{equation}
where $E_S$ approximates the true ground-state energy $E_0$ and $\lvert\psi_S\rangle = \sum_{s\in S} c_s \lvert s\rangle$ is the approximate ground vector in $\mathrm{span}(S)$. This is equivalent to minimizing the Rayleigh quotient in \eqref{eq:restricted_rayleigh}. The primary degree of freedom in this approach is how to choose and expand the subspace $S$.

\section{Problem Statement}
We assume access to:
\begin{itemize}
\item Hamiltonian $H = \sum_\ell c_\ell P_\ell$ as a list of Pauli terms.
\item Finite-shot computational-basis measurement counts from an (imperfect) state preparation.
\end{itemize}
We aim to estimate the true ground-state energy $E_0$ with minimal classical computation and without requiring additional quantum measurements beyond the initial samples (in the basic version studied here).

Let $S_0$ be the initial set of basis states obtained from measurement outcomes, e.g., the top-$K$ most frequent bitstrings. Standard SQD solves the restricted eigenproblem only on $S_0$ and returns $E(S_0)$. However, if the sample omits important basis states (common under finite shots and contamination), $E(S_0)$ can be a poor estimate.

A natural extension is to expand the subspace via Hamiltonian connectivity. Let $\mathcal{N}(s)$ denote basis states connected to $s$ by at least one term in $H$, i.e., $\langle k\lvert H\rvert s\rangle \neq 0$. From current $S$, we can generate a candidate set
\begin{equation}
C(S) = \left(\bigcup_{s\in S} \mathcal{N}(s)\right) \setminus S.
\end{equation}
The key challenge is: \emph{How should we choose a subset of candidates from $C(S)$ to add to $S$ in order to maximize improvement in the ground-energy estimate per added basis state?} Random selection provides a baseline but wastes steps on candidates with negligible energy contribution. We seek a principled, computable acquisition function using only quantities accessible from $(H, S, \lvert\psi_S\rangle)$.

\section{Active SQD via Epstein--Nesbet Perturbation Theory}
To formalize the subspace expansion, let the full Hilbert space be decomposed into a direct sum
\begin{equation}
\mathcal{H} = \mathcal{S} \oplus \mathcal{C},
\end{equation}
where $\mathcal{S} = \mathrm{span}(S)$ is the current subspace and $\mathcal{C}$ is its complement spanned by computational basis vectors not in $S$ (or, in practice, the finite candidate pool $C(S)$). Let $\lvert\psi_S\rangle$ be the normalized lowest-energy eigenvector of $H_S$ with eigenvalue $E_S$. We consider adding a basis state $\lvert k\rangle \in \mathcal{C}$ and ask how much $E_S$ would improve.

Epstein--Nesbet (EN) perturbation theory is commonly used in selected CI to estimate the second-order energy correction from external determinants \cite{epstein1926}. For a single external basis state $\lvert k\rangle$, the EN-inspired contribution takes the form
\begin{equation}
\Delta E_k^{(2)} \approx \frac{\left|\langle k\lvert H\rvert\psi_S\rangle\right|^2}{E_S - H_{kk}},
\label{eq:en}
\end{equation}
where $H_{kk} = \langle k\lvert H\rvert k\rangle$. For ground states, typically $E_S < H_{kk}$ for many $k$, making the denominator negative; thus $\Delta E_k^{(2)}$ is often negative, indicating an energy lowering (improvement). Because we aim to \emph{rank} candidates by their expected impact, we use the magnitude-based acquisition function
\begin{equation}
a(k) = \frac{\left|\langle k\lvert H\rvert\psi_S\rangle\right|^2}{\left|E_S - H_{kk}\right|}.
\end{equation}
If the full candidate space were included without taking absolute values, the signed sum would correspond to the standard Epstein–Nesbet second-order correction. Our acquisition function instead uses a magnitude-based surrogate to rank candidates by expected impact. Selecting candidates with the largest $a(k)$ therefore constitutes a greedy approximation to the optimal subspace expansion that maximally lowers the energy at second order. A small regularization is introduced only to avoid numerical instabilities when $E_S \approx H_{kk}$.

To compute this score efficiently from the restricted solution $\lvert\psi_S\rangle = \sum_{j\in S} c_j \lvert j\rangle$, we evaluate the numerator as
\begin{equation}
\langle k\lvert H\rvert\psi_S\rangle = \sum_{j\in S} c_j \langle k\lvert H\rvert j\rangle.
\label{eq:num}
\end{equation}
Crucially, for Pauli-string Hamiltonians, $\langle k\lvert H\rvert j\rangle$ is sparse: each Pauli term maps a basis state $j$ to exactly one basis state (with a phase and coefficient). Therefore, only a small fraction of pairs $(k,j)$ contribute nonzero values. In practice, we restrict the sum in \eqref{eq:num} to \emph{dominant} components with $|c_j|^2$ above a threshold, which reduces cost and aligns with the intuition that only important configurations should drive exploration.

\section{Algorithm}

\subsection{Overview}
AS-SQD proceeds iteratively:
\begin{enumerate}
\item Initialize $S$ from the top-$K$ most frequent measured bitstrings.
\item Solve the restricted eigenproblem on $S$ to obtain $(E_S,\lvert\psi_S\rangle)$.
\item Generate candidate basis states $C(S)$ connected by $H$.
\item Compute acquisition scores $a(k)$ for $k\in C(S)$.
\item Add the top-$B$ candidates by score to $S$ and repeat.
\end{enumerate}

\subsection{Pseudocode}
The procedure is formalized in Algorithm~\ref{alg:assqd}.

\begin{algorithm}[H]
\caption{Active Sampling Sample-based Quantum Diagonalization (AS-SQD)}
\label{alg:assqd}
\begin{algorithmic}[1]
\REQUIRE Pauli Hamiltonian $H$, measurement counts $\{(b, n_b)\}$, parameters $K,B,T$, thresholds $\tau,\epsilon$
\STATE $S \leftarrow$ top-$K$ bitstrings by count
\FOR{$t=1$ to $T$}
\STATE Build restricted matrix $H_S =[\langle s_i|H|s_j\rangle]$
\STATE Solve $H_S \bm{c} = E_S \bm{c}$ for lowest eigenpair; form $\lvert\psi_S\rangle=\sum_{s\in S} c_s \lvert s\rangle$
\STATE $D \leftarrow \{ s\in S: |c_s|^2 > \tau \}$ \COMMENT{dominant support}
\STATE $C \leftarrow \left(\bigcup_{s\in D} \mathcal{N}(s)\right)\setminus S$ \COMMENT{connected candidates}
\FORALL{$k\in C$}
\STATE $H_{kk} \leftarrow \langle k|H|k\rangle$
\STATE $\nu_k \leftarrow \sum_{s\in D} c_s \langle k|H|s\rangle$
\STATE $a(k) \leftarrow |\nu_k|^2 / \max(|E_S - H_{kk}|,\epsilon)$
\ENDFOR
\STATE Add to $S$ the top-$B$ candidates by $a(k)$
\ENDFOR
\RETURN $E_S$
\end{algorithmic}
\end{algorithm}

\subsection{Baselines}
We compare three methods:
\begin{itemize}
\item \textbf{Standard SQD:} use only $S_0$ from samples; no expansion.
\item \textbf{Random SQD:} iteratively expand by adding a random subset of candidates from $C(S)$ each step.
\item \textbf{AS-SQD (proposed):} expand using the perturbation-guided acquisition function (2).
\end{itemize}

To isolate the components of our acquisition function, we additionally introduce three heuristic ablation baselines:
\begin{itemize}
\item \textbf{Coupling-only:} score candidates by the numerator magnitude $|\langle k|H|\psi_S\rangle|^2$.
\item \textbf{Denom-only:} score candidates by the inverse denominator $1/|E_S - H_{kk}|$.
\item \textbf{Diag-only:} score candidates purely by the lowest diagonal energy $-H_{kk}$.
\end{itemize}

\section{Experimental Setup}
\subsection{Model: Disordered Heisenberg Chain}
We evaluate on a 1D Heisenberg model with periodic boundary conditions:
\begin{equation}
H = J\sum_{i=1}^{n} \left(X_i X_{i+1} + Y_i Y_{i+1} + Z_i Z_{i+1}\right)
+ \sum_{i=1}^{n} h_i Z_i,
\label{eq:heisenberg}
\end{equation}
with $J=1$ and random longitudinal fields $h_i \sim \mathcal{N}(0,h^2)$ (we use $h=0.5$). This Hamiltonian is a standard benchmark for many-body algorithms and is naturally expressed in Pauli terms. 
For classical contamination experiments, we test $n\in\{8,10,12,16\}$ qubits, where the Hilbert dimension is $2^n$ (up to 65{,}536 at $n=16$). 
For hardware experiments, system size is limited to $n\leq12$ due to device constraints.

We also extend our validation to the 1D Transverse-Field Ising Model (TFIM) to ensure generality across Hamiltonian structures:
\begin{equation}
H = -J\sum_{i=1}^{n} Z_i Z_{i+1} - h_x\sum_{i=1}^{n} X_i + \sum_{i=1}^{n} g_i Z_i,
\end{equation}
with $J=1$, $h_x=1$, and longitudinal disorder $g_i \sim \mathcal{N}(0,0.5^2)$.

\subsection{Imperfect Preparation and Finite Shots}
To simulate realistic NISQ limitations, we compute (for benchmarking only) the ground and first excited eigenstates $(E_0,\lvert\psi_0\rangle)$ and $(E_1,\lvert\psi_1\rangle)$ of the full Hamiltonian and then generate measurement outcomes from the contaminated distribution \eqref{eq:contamination} with $\eta=0.2$ (i.e., 80\% ground state and 20\% first excited state). We use $N_{\mathrm{shots}}=2000$ (or 3000 for scaled models) samples to form the initial subspace.

Importantly, the solver does not use the eigenvectors; it only receives the sampled bitstrings and the Pauli-term description of $H$.

Furthermore, to test the algorithm against real physical errors, we generated samples directly from an actual IBM Quantum backend (\texttt{ibmq\_pittsburgh})\cite{ibm2016quantum}. We applied Trotterized state evolution circuits which inherently suffer from SPAM errors, gate infidelities, and decoherence, resulting in highly distorted empirical bitstring distributions.

\subsection{Protocol and Metrics}
For each system size $n$, we generate 5 random field instances (seeds) and report the median absolute energy error:
\begin{equation}
\mathrm{Err} = |E_{\mathrm{est}} - E_0|.
\end{equation}
Initialization chooses the top $K=50$ observed bitstrings. Expansion runs for $T=10$ iterations, adding up to $B=20$ basis states per iteration for Random SQD and AS-SQD. Thus the final subspace size is at most $50+10\times 20 = 250$ (often less due to duplicates and limited candidates).

\section{Results}
\subsection{Performance and Scaling under Contaminated Sampling}
\begin{figure}[!tb]
\centering
\includegraphics[width=\linewidth]{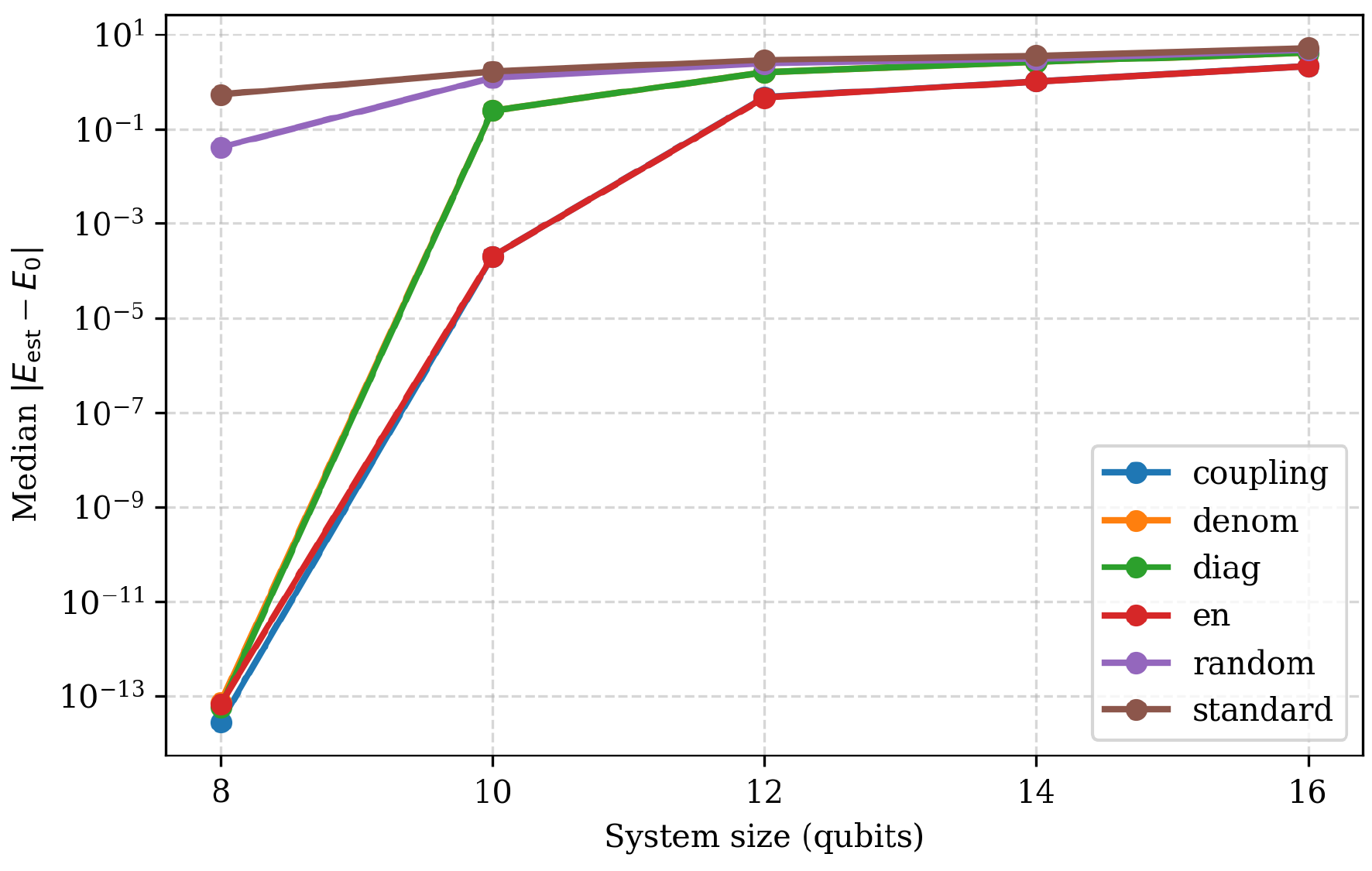}
\caption{Energy error vs. system size for the Heisenberg model under exact contaminated sampling (median over 5 disorder instances).}
\label{fig:scaling_heisenberg}
\end{figure}

\begin{figure}[!tb]
\centering
\includegraphics[width=\linewidth]{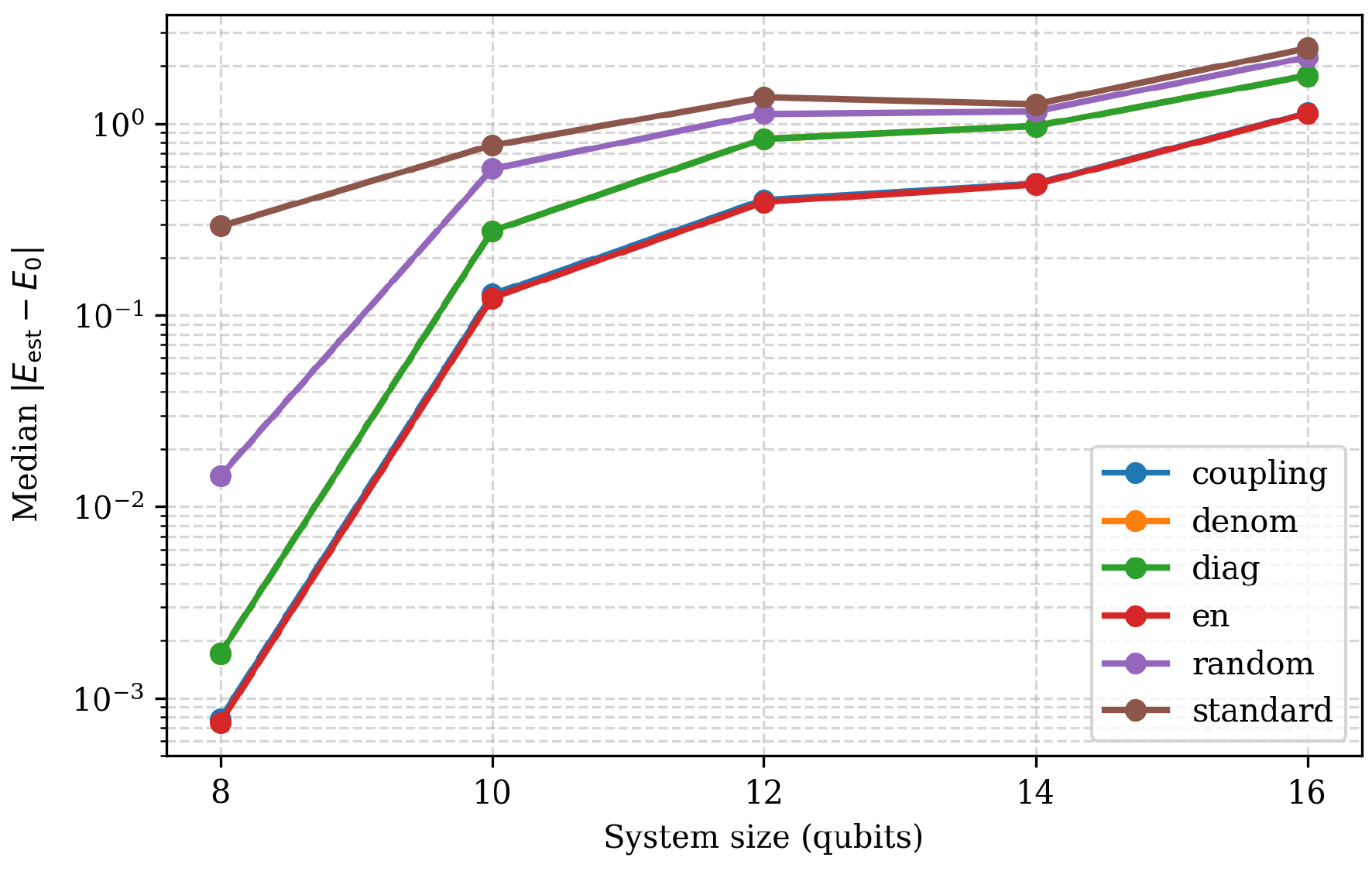}
\caption{Energy error vs. system size for the TFIM model under exact contaminated sampling (median over 5 disorder instances).}
\label{fig:scaling_tfim}
\end{figure}

To evaluate the algorithmic performance and scalability, we first tested the models under the exact contamination sampling regime using 3000 measurement shots. Fig.~\ref{fig:scaling_heisenberg} and Fig.~\ref{fig:scaling_tfim} illustrate the median absolute energy error as a function of system size $n$ for the Heisenberg and TFIM models, respectively.

For small systems like $n=8$ (Hilbert space dimension 256), random expansion effectively reaches a sufficiently expressive subspace within the iteration budget; both Random SQD and AS-SQD attain near machine-precision agreement with the exact ground-state energy $E_0$. However, as the system size increases to $n=10$ and beyond, the performance gap between approaches becomes highly pronounced. Standard SQD errors increase rapidly with $n$ because the empirical samples increasingly miss important off-diagonal support. Random expansion exhibits limited improvement at larger $n$ within a small iteration budget, indicating poor sample efficiency in exploring the combinatorially growing connected basis graph.

Conversely, AS-SQD consistently achieves the best accuracy across all evaluated sizes. Both the full Epstein--Nesbet score (denoted as `en') and the coupling-only heuristic efficiently navigate the basis graph, identifying high-impact states and maintaining remarkably low median errors even at $n=16$ (a Hilbert space dimension of 65,536). This confirms that perturbation-guided basis acquisition effectively concentrates computation on energetically relevant directions, successfully bypassing the exponential bottleneck of blind subspace expansion.

\subsection{Real Hardware Validation on IBM Quantum}
\begin{figure}[!tb]
\centering
\includegraphics[width=\linewidth]{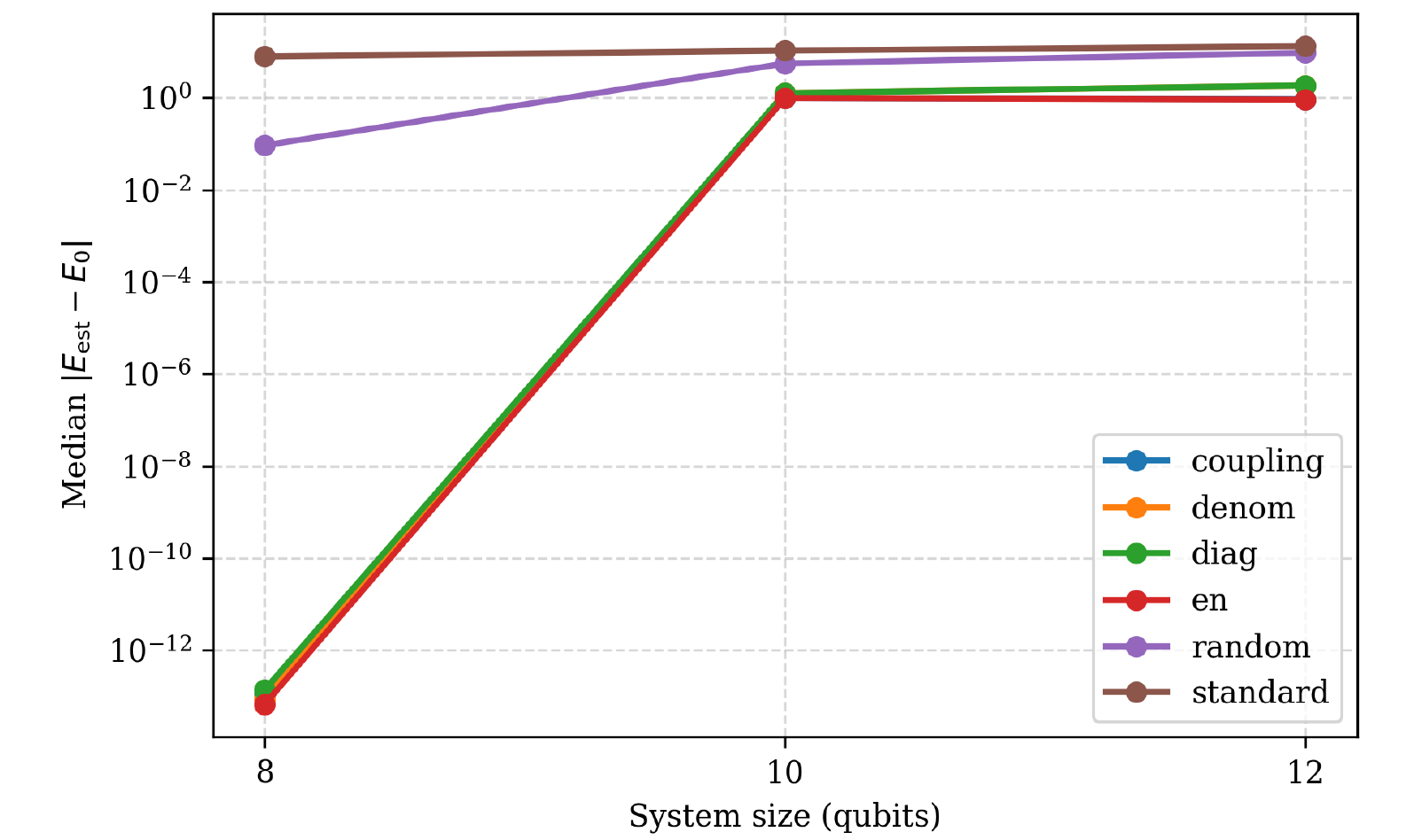}
\caption{Energy error vs. system size using physical samples from IBM Quantum (\texttt{ibmq\_pittsburgh}). Note the remarkable recovery at $n=8$ where AS-SQD successfully filters hardware noise to identify the exact ground state subspace.}
\label{fig:scaling_hardware}
\end{figure}

To test the method against severe physical errors, we evaluated AS-SQD using bitstrings sampled from the \texttt{ibmq\_pittsburgh} quantum processor. Fig.~\ref{fig:scaling_hardware} illustrates the median errors across different system sizes for the Heisenberg model. Despite extreme SPAM and gate noise populating irrelevant computational basis states in the initial distribution, AS-SQD at $n=8$ identified and converged precisely to the true ground state (error $\sim 10^{-14}$). Across all tested hardware sizes, AS-SQD outperforms the standard baselines by a substantial margin, demonstrating strong robustness against physical noise.

\subsection{Ablation Study}
\begin{figure}[!tb]
\centering
\includegraphics[width=\linewidth]{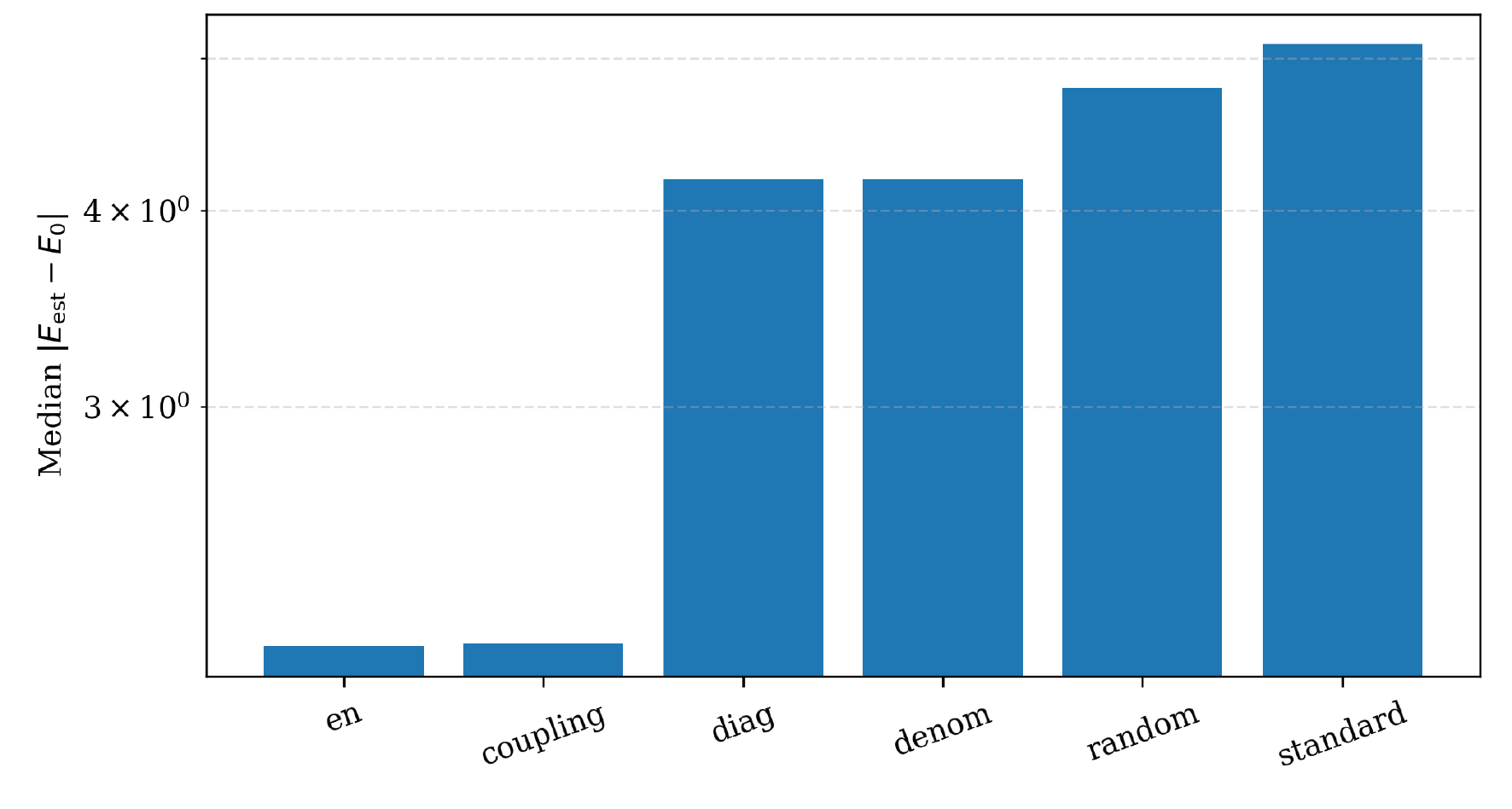}
\caption{Ablation study of acquisition functions at $n=16$ (Heisenberg model under exact contaminated sampling). The full perturbation score (en) and coupling-only significantly outperform energy-based heuristics (denom, diag) and standard baselines.}
\label{fig:ablation}
\end{figure}

To dissect the mechanics of the acquisition function, we compared AS-SQD (`en') to the isolated heuristic scores at $n=16$ (Fig.~\ref{fig:ablation}). Scoring solely by diagonal energies (`diag') or energetic proximity (`denom') performs poorly compared to the full score. In contrast, the `coupling-only' heuristic matches the performance of the full Epstein-Nesbet score. This demonstrates that the matrix element magnitude $|\langle k|H|\psi_S\rangle|^2$ acts as the dominant physical compass in Hilbert space, and the full EN score optimally balances it with energetic proximity.

\subsection{Effect of Candidate Proposal Horizon}
\begin{figure}[!tb]
\centering
\includegraphics[width=\linewidth]{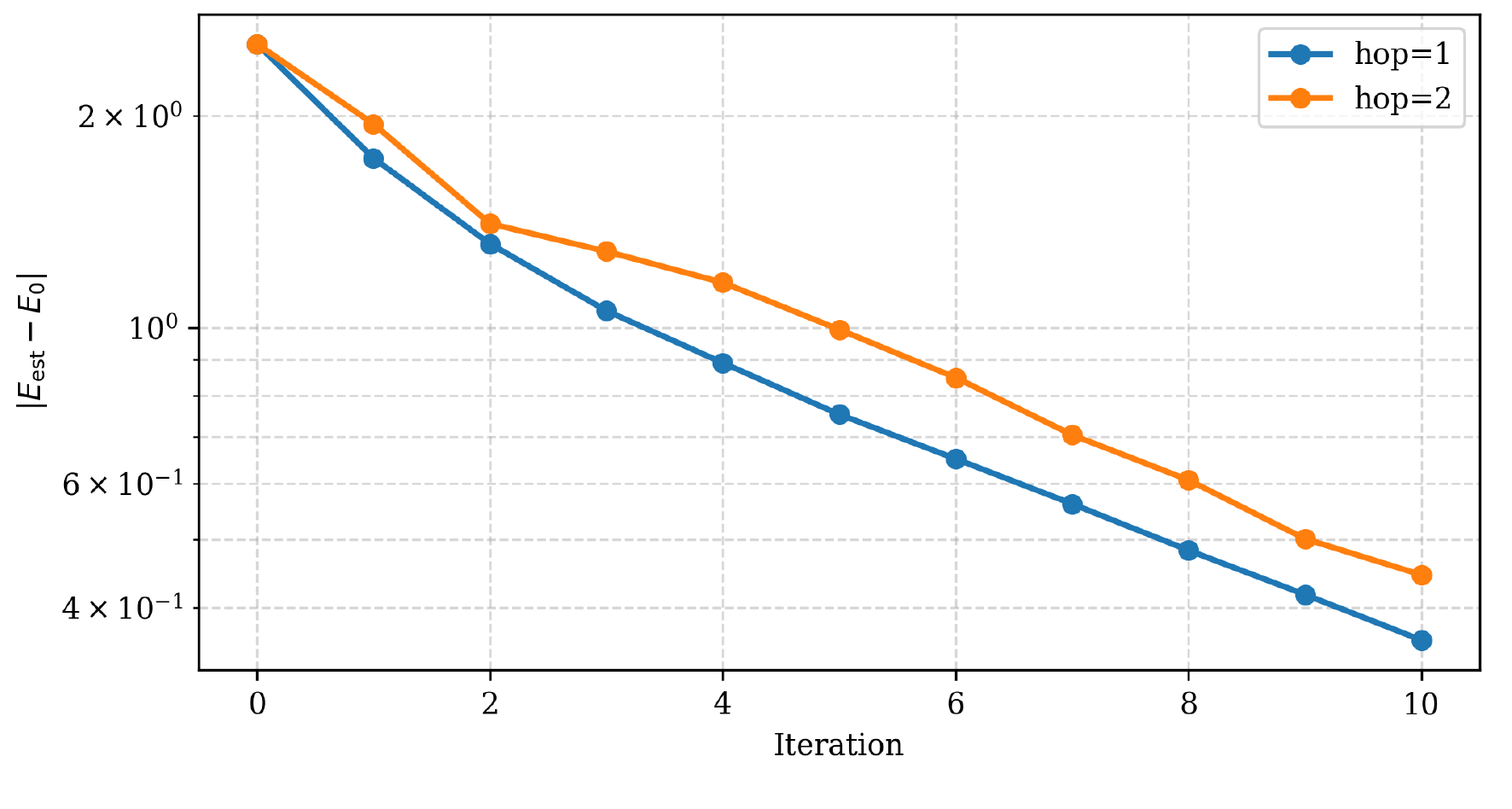}
\caption{Representative error trace ($n=12$ Heisenberg model) demonstrating that extending candidate proposals to 2-hops slows down convergence compared to the standard 1-hop approach. The 1-hop connectivity effectively acts as a strong inductive bias under a fixed addition budget.}
\label{fig:hop_trace}
\end{figure}

To explore the limits of Hamiltonian connectivity, we tested generating candidates up to 2-hops away from the subspace support using a multi-step heuristic score. Interestingly, as shown in Fig.~\ref{fig:hop_trace}, expanding the search horizon to 2-hops actually slowed down convergence under a fixed addition budget ($B=20$). This reveals that the strict 1-hop restriction acts as a highly beneficial inductive bias, preventing the candidate pool from growing excessively and forcing the algorithm to greedily exploit the most immediate and energetically relevant neighborhood.

\section{Discussion}
Standard SQD fails under finite shots because the empirical sample often omits basis states with small probability, while excited-state contamination further displaces important ground-state components from the initial subspace. Expanding this subspace via random connected states is highly inefficient; as system size and interaction complexity increase, an uninformed random selection quickly wastes its limited budget on low-impact states. AS-SQD overcomes these issues by using an acquisition score that simultaneously evaluates coupling strength ($|\langle k\lvert H\rvert\psi_S\rangle|^2$) and energetic proximity ($|E_S-H_{kk}|$). As confirmed by our ablation study, prioritizing strongly connected candidates is vital for efficiently navigating the combinatorial basis graph. This perturbation-guided approach also provides inherent robustness to real-world hardware noise. Physical measurement errors and bit-flips yield observed bitstrings with near-zero off-diagonal couplings to the true low-energy manifold and very high diagonal energies. Consequently, the acquisition function naturally assigns near-zero scores to bitstrings affected by SPAM or gate errors, intrinsically filtering out noise without requiring classical error mitigation techniques such as zero-noise extrapolation or probabilistic error cancellation \cite{kandala2019errormitigation, temme2017errormitigation, endo2021errormitigation}.

Regarding computational complexity, building and diagonalizing the restricted matrix $H_S$ takes $O(m^3)$ classically, where $m=|S|$ is the subspace size. Because $m$ is entirely decoupled from the full Hilbert space dimension $2^n$ and kept deliberately small, the algorithm is highly scalable. Candidate scoring scales linearly with the number of Pauli strings rather than exponentially, successfully bypassing combinatorial bottlenecks. 

While AS-SQD demonstrates significant advantages, we note certain limitations: our current benchmarking relies on classical exact diagonalization for data generation, and numerical instabilities when $E_S \approx H_{kk}$ require small regularizations. Looking forward, framing AS-SQD as a physics-informed active learning loop \cite{settles2009active} in Hilbert space naturally suggests several promising extensions. Because perturbation theory supplies a computationally cheap proxy objective aligned with energy minimization, future work could explore adaptive quantum measurement allocation or bandit-style expansion budgets to further optimize sample efficiency under strict hardware constraints.

\section{Conclusion}
We presented \emph{Active Sampling Sample-based Quantum Diagonalization (AS-SQD)}, a perturbation-guided method for selecting which computational basis states to add to a sample-derived subspace in order to efficiently estimate ground-state energies from finite-shot quantum measurements. AS-SQD reframes SQD as an active learning problem under NISQ constraints and introduces a practical acquisition function based on Epstein--Nesbet second-order energy corrections.

In experiments on Heisenberg and TFIM chains (up to 16 qubits) as well as IBM Quantum hardware, AS-SQD outperformed standard SQD and random expansion, demonstrating improved sample efficiency and robustness against SPAM and gate errors. Future work includes adaptive measurement allocation, enhanced acquisition functions, and evaluation on molecular electronic structure Hamiltonians. More broadly, AS-SQD illustrates a general principle for quantum AI in the NISQ era: \emph{use physics-derived acquisition functions to decide what information to incorporate next under strict sampling constraints}.

\section*{Acknowledgment}
Part of the results of this research were obtained with support from the ``NEDO Challenge, Quantum Computing \textit{Solve Social Issues !}'' by New Energy and Industrial Technology Development Organization (NEDO).

\section*{Data Availability}
All parameters used in the simulations are described in the Methods section and are also available in the provided code.
All program code is freely available on GitHub at \url{https://github.com/G0wOz1PV/SQD}.
The software supports systems running Python 3.7 or later.

\bibliography{main}

\end{document}